\documentclass{article}
\usepackage{a4wide,epsfig,url}

\DeclareSymbolFont{AMSb}{U}{msb}{m}{n}
\DeclareSymbolFontAlphabet{\mathbb}{AMSb}

\newcommand{\ket}[1]{|#1\rangle}
\newcommand{\classical}[1]{\ket{#1}}
\newcommand{\mean}[1]{\overline{#1}}
\newcommand{\order}[1]{{\cal O}(#1)}
\newcommand{\norm}[1]{|\!|#1|\!|}

\newcommand{\complex}{{\mathbb C}}

\title{Reasoning about Grover's Quantum Search Algorithm using  
Probabilistic \textit{wp}}

\author{Michael Butler and Pieter Hartel\\
	Department of Electronics \& Computer Science\\
	University of Southampton\\
	 Southampton SO17 1BJ\\
	 United Kingdom\\
	\url{ {mjb,phh}@ecs.soton.ac.uk} }

\date{8 September, 1998}

\begin{document}

\maketitle

\begin{abstract}
Grover's search algorithm is designed to be executed on a quantum 
mechanical computer.  In this paper, the probabilistic 
\textit{wp}-calculus is used to model and reason about Grover's 
algorithm.  It is demonstrated that the calculus provides a 
rigorous programming notation for modelling this and other quantum 
algorithms and that it also provides a 
 systematic framework of analysing such algorithms.

\end{abstract}

\section{Introduction}  \label{sec:intro}

Quantum computers are a proposed means of using quantum mechanical 
effects to achieve efficient computation.  Quantum mechanical systems 
may be in superpositions of several different states simultaneously.
The central idea of quantum computers is to perform operations on 
these superposed states simultaneously and thus achieve a form of 
parallel computation.  These devices were proposed in the early 
1980's \cite{Benioff80,Deutsch85}.

One essential phenomenon of quantum mechanics is that the measurement 
of a superposed system forces it into a single classical state.  Each 
superposed state is present with a certain amplitude and an 
observation causes it to collapse to that state with a probability 
that depends on its amplitude.  This means that, although many 
computations may be performed in parallel on a quantum device, the 
result of only one of these may be observed.  This may seem like a 
severe limitation, but several ingenious algorithms have been devised 
which work by increasing the amplitude of the desired outcome before 
any observation is performed and thus increasing the likelihood of the 
observed outcome being the desired one.

One such algorithm is Grover's quantum search algorithm 
\cite{Grover97} which performs 
a search on an unstructured search space of size $N$ in 
$\order{\sqrt{N}}$ steps.  
To find the desired search value with 100\% probability
in such a space,
a classical computer cannot do better than a linear time search.
Grover's algorithm performs 
operations on a superposition of all possible search values that serve 
to increase the amplitude of the desired search value.  Grover shows 
that within $\order{\sqrt{N}}$ steps there is a greater than 50\% 
chance of finding the desired search value.  
\cite{BoyerEtAl98} proved a stronger result for the algorithm 
showing that the correct search value can be found in  
$\order{\sqrt{N}}$ with almost 100\% probability.

In this paper, we apply the probabilistic weakest-precondition 
(\textit{wp}) calculus of \cite{MorganEtAl96} to Grover's 
algorithm to redevelop the result of \cite{BoyerEtAl98} in a more systematic 
way.  The probabilistic \textit{wp}-calculus is an extension of 
Dijkstra's standard \textit{wp}-calculus \cite{Dijkstra76} developed 
for reasoning about the correctness of imperative programs.  The 
extension supports reasoning about programs containing 
probabilistic choice.  The measurement of a quantum superposition is 
an example of a probabilistic choice.

Use of the probabilistic \textit{wp}-calculus contributes two essential 
ingredients to the analysis of quantum algorithms.  Firstly it 
provides an elegant and rigorous programming language for describing 
quantum algorithms.  The existing literature uses block diagrams and 
structured English which can be cumbersome and potentially ambiguous.  
Secondly, the probabilistic \textit{wp}-calculus provides a set of rules for 
the systematic analysis of the correctness of algorithms.  In the case 
of standard algorithms, the calculus is used to determine whether a 
program achieves some desired outcome.  In the case of probabilistic 
algorithms, the calculus is used to reason about the probability of a 
program achieving some desired outcome.

This paper is not simply about re-presenting a known result about Grover's 
algorithm but it also aims to demonstrate that the probabilistic 
\textit{wp}-calculus is suitable for both modelling and reasoning about a quantum 
algorithm.  Boyer et al have already derived the same result that we 
derive here but they do so in a less systematic way.  Our hope is 
that the approach used here could be applied fruitfully to other 
quantum algorithms and may even aid the development of new quantum 
algorithms.

The paper is organised as follows.  In Section~\ref{sec:qubit}, we 
give a sufficient overview of quantum theory.  In 
Section~\ref{sec:model}, we present our approach to modelling quantum 
computation using the programming language of the probabilistic 
\textit{wp}-calculus.  In Section~\ref{sec:grover}, we present 
Grover's algorithm using the approach of Section~\ref{sec:model}.
In Section~\ref{sec:pwp}, we give a sufficient 
introduction to the rules of the probabilistic \textit{wp}-calculus 
and in Section~\ref{sec:reason}, we use the \textit{wp}-calculus to 
derive a formula for the probability of success of Grover's algorithm.

\section{Quantum Systems and Qubits}  \label{sec:qubit}

%\subsection{Two State Systems}

In quantum mechanics, a superposition of two states $A$ and $B$ is 
represented in Dirac's notation as follows:
\begin{eqnarray*}
	S &=& \alpha\ket{A} ~+~ \beta\ket{B}.
\end{eqnarray*}
System $S$ is said to be in a superposition of $A$ and $B$.
$\ket{A}$ and $\ket{B}$ are the \textit{basis states} and $\alpha$ and 
$\beta$ are \textit{amplitudes}.  The amplitudes may be complex numbers.

Let $\norm{z}^{2}$ be the square norm\footnote{The square norm of any
complex number
$a+bj$ is $a^{2}+b^{2}$.} of complex number $z$.
Observation of $S$ will cause the system to collapse to state
 $\ket{A}$ with probability $\norm{\alpha}^{2}$ 
and to $\ket{B}$ with probability $\norm{\beta}^{2}$.
The  probabilities must sum to 1:
\begin{eqnarray*}
	\norm{\alpha}^{2} ~+~ \norm{\beta}^{2} &=& 1.
\end{eqnarray*}

A \textit{qubit} is a two state quantum system in which the basis states 
are labelled 0 and 1:
\begin{eqnarray*}
	S &=& \alpha\ket{0} ~+~ \beta\ket{1}.
\end{eqnarray*}
A classical bit has $\alpha=1$ and $\beta=0$  or  
$\alpha=0$ and $\beta=1$.

A qubit evolves from one superposition to another using a 
quantum gate (or function) $F$:
\begin{eqnarray*}
	F(~ \alpha\ket{0} ~+~ \beta\ket{1} ~)  &=& 
	  \alpha'\ket{0} ~+~ \beta'\ket{1}.
\end{eqnarray*}
$F$ must be \textit{unitary} which means that
\begin{itemize}
\item probabilities are preserved: 
        $\norm{\alpha'}^{2} ~+~ \norm{\beta'}^{2} ~~=~~
         \norm{\alpha}^{2} ~+~ \norm{\beta}^{2}$, and
\item $F$ has an inverse.
\end{itemize}

In quantum mechanics, a transformation $F$ is usually modelled using matrix
multiplication:
\begin{eqnarray*}
 F(~ \alpha\ket{0} ~+~ \beta\ket{1} ~)  &=& 
 U_{F} \times \left(\begin{array}{l}
                 \alpha \\ \beta
                \end{array}
          \right),
\end{eqnarray*}
where $U_{F}$ is a $2\times 2$ \textit{unitary} matrix.
Matrix $U$ is unitary if $U.U^{\dagger} = U^{\dagger}.U = I$ where
$U^{\dagger}$ is the conjugate transpose of $U$.  It can be shown that
such a transformation defined by unitary matrix $U_{F}$ is unitary
\cite{BernsteinVazirani93}.

A quantum superposition may have an arbitrary number of basis states,
 not just two.  An $N$-state superposition is represented as:
\begin{eqnarray*}
	S &=& \sum_{i=0}^{N-1} \alpha_{i} \ket{i} .
\end{eqnarray*}

Observation of $S$ will cause it to
collapse to state $\ket{i}$ with probability $\norm{\alpha_{i}}^{2}$.
Again, the probabilities must sum to 1:
\begin{eqnarray*}
	\sum_{i=0}^{N-1} \norm{\alpha_{i}}^{2}  &=& 1 .
\end{eqnarray*}

A \textit{quantum register} is a collection of qubits and an 
$L$-qubit register gives rise to a system with $2^{L}$ basis states.
Like qubits, quantum registers evolve under unitary transformations.

For further details on quantum computation, the reader is referred to
papers such as \cite{Berthiaume,Ekert}.

\section{Modelling Quantum Computers}  \label{sec:model}

A quantum computer is a collection of quantum registers and quantum 
gates.  In this section, we introduce ways of modelling various 
aspects of quantum computation using the programming language of the 
probabilistic \textit{wp}-calculus.   
We use a subset of the language which 
includes standard assignment, probabilistic assignment, sequential 
composition and simple loops.

%\subsection{Modelling Superpositions}

Firstly, we model an $N$-state quantum system as a function from
state indices to complex numbers:
$ S: (0..N-1) \rightarrow \complex $.

A superposition of the form 
\begin{eqnarray*}
	\sum_{i=0}^{N-1} \alpha_{i} \ket{i} 
\end{eqnarray*}
is modelled  by the function $S$ where for $0\leq i < N$:
\begin{eqnarray*}
	S(i)  &=& \alpha_{i}.
\end{eqnarray*}

A classical state $i$ is modelled by the function which is zero
everywhere except at $i$ which we write as $\classical{i}$:
\begin{eqnarray*}
	\classical{i}(j)  &=& 1,~ \hbox{if}~ i=j\\
	&=& 0,~ \hbox{otherwise}.
\end{eqnarray*}

Transformation of a quantum state is modelled by a standard 
\textit{assignment} statement:
\begin{eqnarray*}
	S ~:=~ F(S).
\end{eqnarray*}
 $F$ must be unitary for this to be a valid quantum transformation.

We shall find it convenient to use  lambda abstraction
 to represent 
transformations:
$(\lambda i ~|~ 0\leq i<N \cdot E)$ represents the function that takes 
an argument $i$ in the range $0..N-1$ and returns the value $E$.
For example, the unitary transformation that inverts 
the amplitude of each basis state is modelled as follows:
\begin{eqnarray*}
	S ~:=~ (~\lambda i ~|~ 0\leq i<N \cdot -S(i)~) .
\end{eqnarray*}

Sequencing of transformations is modelled using \textit{sequential 
composition}: let $T_{1}$ and $T_{2}$ be transformations, then their
sequential composition is written $T_{1};T_{2}$.

The loop which iterates $C$ times over a transformation $T$ is 
written  %\linebreak
${\bf do}~ C ~{\bf times}~ T ~{\bf od}$.

%\subsection{Modelling Observation}

We model the observation of a quantum system using a 
\textit{probabilistic assignment} statement.  
In the simple case, this is a statement of the form:
\begin{eqnarray*}
	x &:=&  E ~~@~~ p, \\
	  &  &  F ~~@~~ (1-p).
\end{eqnarray*}
This says that $x$ takes the value $E$ with probability $p$ and
the value $F$ with probability $1-p$.  For example, a coin flip is
modelled by 
\begin{eqnarray*}
	coin &:=&  head ~~@~~ 0.5, \\
	  &  &  tail ~~@~~  0.5.
\end{eqnarray*}

Observation of a two state superposition
  forces the system into a classical state.  This is 
modelled with the following probabilistic assignment:
\begin{eqnarray*}
	S &:=&  \classical{0} ~~@~~ \norm{S(0)}^{2}, \\
	  &  &  \classical{1} ~~@~~ \norm{S(1)}^{2}.
\end{eqnarray*}

A generalised probabilistic statement has the form
\begin{eqnarray*}
	x &:=&  E_{i} ~~@~~ p_{i},~~~  0\leq i < N,
\end{eqnarray*}
where $(\sum_{i=0}^{N-1} p_{i}) = 1$.

Now observation of an $N$-state quantum system $S$ may be modelled by
\begin{eqnarray*}
	S &:=&  \classical{i} ~~@~~ \norm{S(i)}^{2},~~~  0\leq i < N.
\end{eqnarray*}
That is, $S$ collapses to the
 classical state $i$ with probability $\norm{S(i)}^{2}$.

\section{Grover's Search Algorithm}  \label{sec:grover}

The Grover search problem may be stated as follows:
\begin{quotation}
\noindent
  Given a 
 function $f:(0..N-1)\rightarrow \{0,1\}$ 
that is zero everywhere except for one argument
 $x_{0}$, where  
 $ f(x_{0}) =1$, find
  that argument $x_{0}$.
\end{quotation}
  
The algorithm makes use of the mean of a superposition $S$, written 
$\mean{S}$, where
\begin{eqnarray*}
   \mean{S} &=& \frac{\sum_{i=0}^{N-1}S(i)}{N}.
\end{eqnarray*}

\begin{figure}
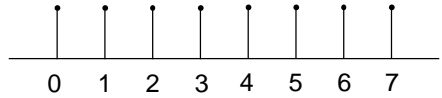

\begin{eqnarray*}
\begin{array}{ll}
	S ~:=~ (\lambda i ~|~ 0\leq i < N \cdot \frac{1}{\sqrt{N}} ) ~; 
	&  \hbox{ ~~~\textit{Init} } \\[1.5ex]
	{\bf do}~ C ~{\bf times}  \\[1ex]
	~~~~ \begin{array}{l}
	     S ~:=~ (\lambda i ~|~ 0\leq i < N \cdot S(i) - 2.f(i).S(i)~) ~; \\[1ex]
	     S ~:=~ (\lambda i ~|~ 0\leq i < N \cdot 2.\mean{S} - S(i)~)  
	     \end{array}
	 & \hbox{ ~~~\textit{Body} }\\[1ex]
	{\bf od} ~; \\[1.5ex]
	S ~:=~ \classical{i} ~~@~~ \norm{S(i)}^{2},~~~~ 0\leq i < N
	 & \hbox{ ~~~\textit{Measure} }
\end{array}
\end{eqnarray*}
\caption{Grover's search algorithm.} \label{fig:grover}
\end{figure}

The algorithm is represented in the  programming language of the 
probabilistic \textit{wp}-calculus in Fig.~\ref{fig:grover}.
The initialisation of this algorithm sets the system $S$ up in an equal 
superposition of all possible basis states.  Successive iterations of 
the loop then serve to increase the amplitude of the search argument 
$x_{0}$  while decreasing the amplitude of the other arguments.
To see why this is so, consider the case of $N=8$. 
The initialisation sets $S$ up in an equal superposition of the eight 
possible states, represented diagrammatically as follows:\\

\newlength{\grover}
\setlength{\grover}{0.5\textwidth}
\addtolength{\grover}{-3cm}

\noindent\hspace{\grover}
 \epsfig{file=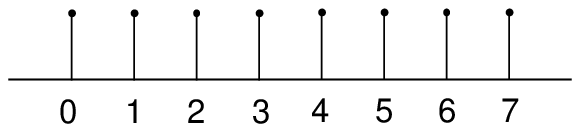}\\

The first step of the loop body replaces each $S(i)$ with $S(i)-2.f(i).S(i)$.
This inverts $S(i)$ about the origin in the case 
that $f(i)=1$ and leaves $S(i)$ unchanged in the case that $f(i)=0$.
Assuming that $f(4)=1$, this replaces our example superposition with\\

\noindent\hspace{\grover}
 \epsfig{file=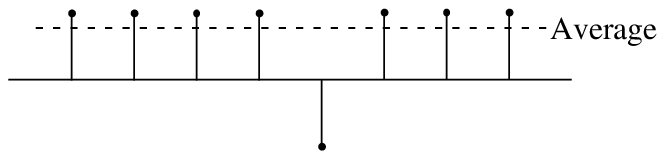}\\

The second step of the loop body inverts each amplitude about the 
average of all the amplitudes resulting in:\\

\noindent\hspace{\grover}
 \epsfig{file=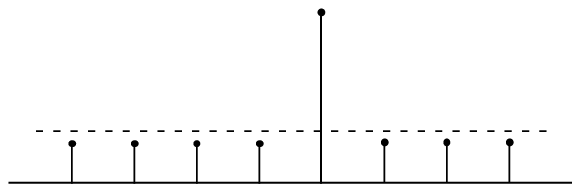}\\

The amplitude of state $\ket{4}$ has increased as a result of the
two steps of the loop body, while the amplitude 
of the others has decreased.

After an optimum number of iterations, $C$, the amplitude of 
$\ket{x_{0}}$ approaches 1 while the amplitude of the other states
approaches 0.
An observation is then performed.  Since the 
amplitude of $\ket{x_{0}}$ approaches 1, the probability of the observation 
yielding $\classical{x_{0}}$ is close to 1.
$C$ depends on the number of states $N$ and, as discussed in the next 
section, it is $\order{\sqrt{N}}$.

\section{Probabilistic \textit{wp}}  \label{sec:pwp}

In two-valued logic, a \textit{predicate} may be modelled as a function
from some state space to the set $\{0,1\}$.
For example,  the predicate $x>y$ evaluates to $1$ in a state in 
which $x$ is greater then $y$ and evaluates to $0$ in any other 
state.  A \textit{probabilistic predicate} generalises the range to the 
continuous space between $0$ and $1$ \cite{MorganEtAl96}.
For example, the probabilistic predicate 
$0.5 \times (x>y)$ evaluates to $0.5$ in a state in 
which $x$ is greater then $y$ and evaluates to $0$ in any other 
state.

In the standard \textit{wp}-calculus, the semantics of imperative programs is 
given using weakest-precondition formulae: for program $prog$ and 
postcondition $post$, %\linebreak 
$wp(prog,post)$ represents the weakest precondition (or maximal set of 
initial states) from which $prog$ is guaranteed to terminate and 
result in a state satisfying $post$.

The \textit{wp} rule for assignment is given by:
\begin{eqnarray}
	wp(~ x:=E,~ post ~)  &=& post[x/E] .   \label{wp:assign}
\end{eqnarray}
Here, $post[x/E]$ represents predicate $post$ with all free 
occurrences of $x$ replaced by $E$.
For example,  
\begin{eqnarray*}
  wp(x:=7,~ x>y) 
  &=& x>y~[x/7]  \\
  &=& 7>y.
\end{eqnarray*}
That is, the assignment $x:=7$ is 
guaranteed to establish $x>y$ provided $7>y$ initially.

The \textit{wp} rule for sequential composition is given by:
\begin{eqnarray} \label{wp:seq}
	wp(~ prog1;prog2,~ post ~)  &=&  wp(~ prog1,~ wp(prog2,~ post ~) ~).
\end{eqnarray}

Both of these rules also apply in the probabilistic 
\textit{wp}-calculus.  The \textit{wp} rule for simple probabilistic 
assignment \cite{MorganEtAl96} 
is given by:
\begin{eqnarray} \label{wp:prob2}
	\lefteqn{wp(~ x~:=~ E~@~p,~ F~@~1-p~ ,~~ post ~)  ~= } \nonumber\\
	 &&  p\times post [x/E] ~+~ (1-p) \times post[x/F].
\end{eqnarray}

In the case of non-probabilistic $post$, $wp(prog,post)$ 
represents  the probability that program $prog$ establishes 
$post$.
For example
\begin{eqnarray*}
&&	wp(~ coin~:=~ head~@~0.5,~ tail~@~0.5,~~ coin=head ~) \\
&=&  ~~~~~~\hbox{by (\ref{wp:prob2})} \\ &&
     0.5\times( coin=head~[coin/head] ) ~+~ 
     0.5\times( coin=head~[coin/tail] )  \\
&=&  ~~~~~~\hbox{substitution} \\ &&
     0.5\times( head=head ) ~+~ 
     0.5\times( tail=head )  \\
&=&  0.5\times 1 ~+~ 
     0.5\times 0 \\
&=&  0.5.
\end{eqnarray*}
That is, a coin flip establishes $coin=head$ with probability $0.5$.

The \textit{wp} rule for the generalised probabilistic assignment is given 
by:
\begin{eqnarray} \label{wp:probn}
	wp(~ x~:=~ E_{i}~@~p_{i},~ 0\leq i <N ,~~ post ~)  &=& 
	 \sum_{i=0}^{N-1} p_{i}\times post [x/E_{i}] .
\end{eqnarray}

The only other programming construct we need in order to model 
Grover's algorithm is the DO-loop.  Since the algorithm only loops a 
constant and finite number of times, we can model 
${\bf do}~C~{\bf times}~prog~{\bf od}$ as a finite sequential 
composition of $C$ copies of $prog$ which we write as
$prog^{C}$.  We have that
\begin{eqnarray}
	prog^{0}  &=& skip     \label{eq:loop0}  \\
	prog^{i+1}  &=&  prog ~;~ prog^{i} . \label{eq:loop1} 
\end{eqnarray}
Here, $skip$ is the statement that does nothing, with
$wp(skip,post)=post$.
The semantics of more general looping constructs is given by least 
fixed points in the usual way, but we do not need that here.

\section{Reasoning about Grover}  \label{sec:reason}

The postcondition we are interested in for the Grover algorithm is 
that the correct solution is found, i.e., $S=\classical{x_{0}}$.
The probability that Grover establishes $S=\classical{x_{0}}$ is given
by $wp(~ Grover,~ S=\classical{x_{0}} ~)$, so we shall calculate this.

The Grover algorithm has the following structure:
\begin{eqnarray*}
\begin{array}{l}
	Init ~; \\
	{\bf do}~ C ~{\bf times}  \\
	~~~~ Body  \\
	{\bf od} ~; \\
	Measure.
\end{array}
\end{eqnarray*}
which we shorten to 
\begin{eqnarray*}
	Init ~;~ 
	Body^{C} ~;~
	Measure.
\end{eqnarray*}

When calculating a formula of the form $wp(~ prog1;prog2,~ post ~)$,
we first calculate \linebreak
$wp(~prog2,~ post ~)$ and then apply $wp(~prog1,\_)$ to 
the result of this.
Thus, to calculate \linebreak
$wp(~ Grover,~ S=\classical{x_{0}} ~)$, we
first calculate $wp(~ Measure,~ S=\classical{x_{0}} ~)$:
\begin{eqnarray}    
	&& wp(~ Measure,~ S = \classical{x_{0}} ~)  \nonumber\\
	&=&  wp(~ S:=\classical{i} ~@~ \norm{S(i)}^2,~ 0\leq i <N,~~~ 
	           S = \classical{x_{0}} ~)  \nonumber\\
	&=&  ~~~~~~ \hbox{ by (\ref{wp:probn}) } \nonumber\\ &&
	      \sum_{i=0}^{N-1} \norm{S(i)}^2 \times 
	         (\classical{i}=\classical{x_{0}}) \nonumber\\
	&=&  ~~~~~~\hbox{ since $( \classical{i} = \classical{x_{0}} )$
	                  is 0 for $i \neq x_{0}$ } \nonumber\\&&
	     \norm{S(x_{0})}^2 .  \label{eq:wpmeasure}
\end{eqnarray}

Next we calculate $wp(~ Body^{C},~  \norm{S(x_{0})}^2 ~)$.
$Body^{C}$ is defined recursively by (\ref{eq:loop0}) and
(\ref{eq:loop1}) so we shall develop recursive equations for
$wp(~ Body^{C},~  \norm{S(x_{0})}^2 ~)$.
First we look at the weakest precondition of a single iteration.
Let $P[S]$ stand for a predicate $P$ containing one or more
free occurrences of variable $S$ and
$P[S']$ stand for $P$ with all free occurrences of $S$ replaced by $S'$.
It is easy to show, using (\ref{wp:assign}) and (\ref{wp:seq}), that
\begin{eqnarray}
	wp(~ Body,~ P[S] ~)  &=&  P[S']   \label{eq:wpbody1} \\
	\hbox{where}~ S'(i)  &=&  
	2.\mean{S} - \frac{4}{N}.S(x_{0}) + (2.f(i) - 1).S(i) .\nonumber
\end{eqnarray}
From (\ref{eq:wpbody1}), we have that
\begin{eqnarray*}
	wp(~ Body,~ \norm{S(x_{0})}^2 ~)  &=& 
	 \norm{S'(x_{0})}^2 \\
	 &=&  \norm{ 2.\mean{S} - \frac{4}{N}.S(x_{0}) + 
	                   (2.f(x_{0}) - 1).S(x_{0}) }^2 \\
	 &=&  \norm{ 2.\mean{S} ~+~ (1 - \frac{4}{N}).S(x_{0}) }^2 .
\end{eqnarray*}
Now this  has the form  $\norm{~  A.\mean{S} ~+~ B.S(x_{0}) ~}^2$
and using (\ref{eq:wpbody1}) we can again show that for any values $A, B$:
\begin{eqnarray}
	wp(~ Body,~ \norm{~ A.\mean{S} ~+~ B.S(x_{0}) ~}^2 ~)  &=& 
	 \norm{~ A'.\mean{S} ~+~ B'.S(x_{0}) ~}^2      \label{eq:wpbody2}\\
	 \hbox{where}~ A' &=&  A + 2.B   \nonumber \\
	 		B' &=&  \frac{ N.B - 2.A - 4.B }{ N } . \nonumber
\end{eqnarray}
This recurring structure suggests that we define $A_{i}$ and 
$B_{i}$ as follows:
\begin{eqnarray}
	 A_{i+1} &=&  A_{i} + 2.B_{i}  \label{eq:A} \\
	 B_{i+1} &=&  \frac{ N.B_{i} - 2.A_{i} - 4.B_{i} }{ N }, \label{eq:B} 
\end{eqnarray}
to give
\begin{eqnarray}
	wp(~ Body,~ \norm{~ A_{i}.\mean{S} ~+~ B_{i}.S(x_{0}) ~}^2 ~)  &=& 
	 \norm{~ A_{i+1}.\mean{S} ~+~ B_{i+1}.S(x_{0}) ~}^2 .
\end{eqnarray}
By induction over $j$, we get
\begin{eqnarray}
	wp(~ Body^{j},~ \norm{~ A_{i}.\mean{S} ~+~ B_{i}.S(x_{0}) ~}^2 ~)  &=& 
	 \norm{~ A_{i+j}.\mean{S} ~+~ B_{i+j}.S(x_{0}) ~}^2 .   \label{eq:wploop}
\end{eqnarray}

We take $A_{0}=0$ and $B_{0}=1$ and apply $Body^{C}$ to 
(\ref{eq:wpmeasure}) as follows:
\begin{eqnarray*}
	&& wp(~ Body^{C},~  \norm{ S(x_{0}) }^2 ~) \\
	&=&    ~~~~~~\hbox{ since $A_{0}=0$, $B_{0}=1$ }\\ &&
	     wp(~ Body^{C},~  
	       \norm{~ A_{0}.\mean{S} ~+~ B_{0}.S(x_{0}) ~}^2 ~)  \\
	&=&   ~~~~~~\hbox{ by (\ref{eq:wploop}) } \\&&
	      \norm{~ A_{C}.\mean{S} ~+~ B_{C}.S(x_{0}) ~}^2 .
\end{eqnarray*}

Finally, we apply the initialisation to this:
\begin{eqnarray*}
	&& wp(~ Init,~  
	        \norm{~ A_{C}.\mean{S} ~+~ B_{C}.S(x_{0}) ~}^2 ~) \\
%	&=& wp(~~ S~:=~(\lambda i ~|~ 0\leq i < N \cdot \frac{1}{\sqrt{N}} ) ,~  
%	        (~ A_{C}.\mean{S} ~+~ B_{C}.S(x_{0}) ~)^2 ~~) \\
	&=&   \norm{~ A_{C}.\frac{1}{\sqrt{N}} ~+~ 
	         B_{C}.\frac{1}{\sqrt{N}} ~}^2 \\
	&=&   \frac{\norm{~ A_{C} + B_{C} ~}^2}{N} .
\end{eqnarray*}

Thus we have shown that:
\begin{eqnarray*}
 wp(~ Grover,~ S=\classical{x_{0}} ~) 
	&=&   \frac{\norm{~ A_{C} + B_{C} ~}^2}{N} .
\end{eqnarray*}
That is, the probability, $P(C,N)$, of observing the correct 
value after $C$ iterations is:
\begin{eqnarray*}
  P(C,N) &=& \frac{\norm{~ A_{C} + B_{C} ~}^2}{N} .
\end{eqnarray*}

Now using standard mathematical analysis techniques, 
we can derive the following closed form for $P(C,N)$:
\begin{eqnarray*}
  P(C,N) &=&  \sin^{2} ((2.C+1).\theta_{N}) \\
  \hbox{where}~ \theta_{N} &=& \arcsin\frac{1}{\sqrt{N}}.
\end{eqnarray*}
This is the same as the formula presented in \cite{BoyerEtAl98}.
The derivation of this closed form is outlined in the appendix.

\begin{figure}
\begin{center}
 \epsfig{file=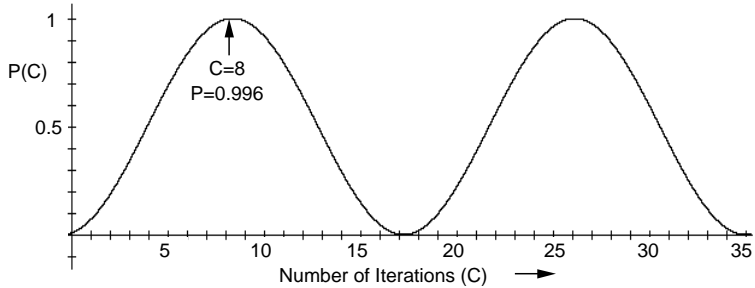,width=4in}
\end{center}
\caption{Probability for Grover search with $N=128$.}
\label{fig:prob}
\end{figure}

It is interesting to note that $P(C,N)$  is periodic in $C$.  This 
can be seen clearly in Fig.~\ref{fig:prob} which graphs $P(C,N)$ 
against $C$ for $N=128$.  Here, an optimum probability of success
is reached 
after eight iterations, where $P(8,128)=0.996$. After eight iterations,
the probability starts to decrease again.  The reason for the decrease 
is that, after eight iterations, 
the average immediately after the inversion about the 
origin operation goes below zero.

\begin{figure}
\begin{center}
 \epsfig{file=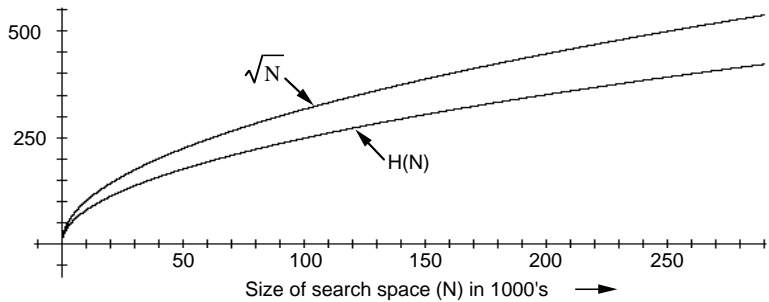,width=4in}
\end{center}
\caption{Optimum number of iterations versus search space size.}
\label{fig:opt}
\end{figure}

We wish to determine the optimum number of iterations for a given $N$.
$P$ reaches a maximum (and a minimum) for a given $N$ when:
\begin{eqnarray*}
	\frac{d}{dx}P(x,N) &=& 0.
\end{eqnarray*}
It is easy to show that the first maximum for a given $N$ is reached at
\begin{eqnarray*}
	\frac{\pi}{4.\theta_{N}} -\frac{1}{2} &~~~& 
	\hbox{where}~ \theta_{N} ~=~ \arcsin\frac{1}{\sqrt{N}}.
\end{eqnarray*}
We call this $H(N)$.  Thus the number of iterations in the Grover 
algorithm for a search space of size $N$
should be the closest whole number to $H(N)$.
In Fig.~\ref{fig:opt}, we graph $H(N)$ and indicate that it is 
$\order{\sqrt{N}}$.

\section{Conclusions}

We have shown how Grover's search algorithm may be represented in the 
programming notation of the probabilistic \textit{wp}-calculus.  Any 
quantum computation consists of unitary transformations and 
probabilistic measurement and these can be modelled in this notation.
Thus any quantum algorithm may be modelled in the notation.  We 
believe that this language provides a more rigorous and elegant means 
of describing quantum algorithms than is normally used in the 
literature.

We have also shown how the rules of the probabilistic 
\textit{wp}-calculus may be used to derive a recursive formula for the 
probability that Grover's algorithm finds the required solution.  
Using standard mathematical techniques, 
we were then able to then find a closed form for this 
probability which corresponds to the formula presented in 
\cite{BoyerEtAl98}.  The \textit{wp}-calculus provides a clear 
and systematic means of stating the required outcome and of
deriving the probability of achieving it.  Of course, it 
does not provide everything for free as we still had to use 
intelligence in recognising the recurring structure and in finding a 
closed form.

In the case of Grover, we were able to derive an exact probability for 
success because the algorithm iterates a fixed number of times.  Some 
algorithms iterate until some condition is met rather a fixed number 
of times.  One such example is a generalisation of Grover's presented 
in \cite{BoyerEtAl98} which deals with the situation where there are 
an unknown number of values $x$ satisfying $f(x)=1$.  In a case like 
this, we need to find the expected number of iterations rather than 
the probability of success.  For future work, we intend to look at how 
these cases may be reasoned about using the probabilistic 
\textit{wp}-calculus.

\appendix

\section{Deriving a Closed Form}

We outline the derivation of the closed form expression for the 
probability of success of Grover's algorithm.
The probability $P(C,N)$ is expressed in terms of the series
$A_i$ and $B_i$, which in turn are defined by the recurrence
equations (\ref{eq:A}) and (\ref{eq:B}). 
To find a closed form for these recurrences we first
compute the generating functions for $A_i$ and $B_i$ using basic
techniques~\cite[Sect. 1.2.9]{Knuth}:

\begin{eqnarray*}
A_{i} ~=~   \frac{2 N i}{N + 2 (2 - N) i + N i^2} & &
B_{i} ~=~ \frac{N - N i}{N + 2 (2 - N) i + N i^2}.
\end{eqnarray*}

Computing the probability involves the sum $A_i+B_i$ and it
seems reasonable to examine the Taylor series expansion of the sum of
the two generating functions.
Assume that $z$ is such that the Taylor series expansion of 
$A_{z}+B_{z}$ converges:

\begin{eqnarray*}
&& A_{z}+B_{z} \\
&=& \frac{N + N z}{N + 2 (2 - N) z + N z^2} \\
&=& ~~~~~~\hbox{ Taylor expansion } \\ &&
    1 + \frac{3 N - 4}{N} z + \frac{5 N^2 - 20 N + 16}{N^2} z^2
                + \frac{7 N^3 - 56 N^2 + 112 N - 64}{N^3} z^3 + \cdots.
\end{eqnarray*}

We now observe that there is a strong similarity between the coefficients
and powers of $N$ in the enumerators above and the coefficients and powers
of $\sin(\theta)$ in the multiple angle formula for $\sin(n \theta)$:

\[
\begin{array}{*{3}{@{~}l}}
\sin(1 \theta)&=& 1 \sin(\theta) \\
\sin(3 \theta)&=& 3 \sin(\theta) -  4 \sin^3(\theta)  \\
\sin(5 \theta)&=& 5 \sin(\theta) - 20 \sin^3(\theta) +  16 \sin^5(\theta)  \\
\sin(7 \theta)&=& 7 \sin(\theta) - 56 \sin^3(\theta) + 112 \sin^5(\theta) - 64 \sin^7(\theta).
\end{array}
\]

This similarity suggests that we express $N$ in the form
$\sin^{-2}(\theta)$ since, for example,
\begin{eqnarray*}
\frac{5 N^2 - 20 N + 16}{N^2}[N/\sin^{-2}(\theta)] &=&
5 - 20 \sin^2(\theta_N) +  16 \sin^4(\theta_N).
\end{eqnarray*}
We write $\theta$ as $\theta_N$ and 
choose it so that $\sin^{-2}(\theta_N)=N$, i.e.,
$\theta_N=\arcsin\frac{1}{\sqrt{N}}$.

Rewriting $N$ as $\sin^{-2}(\theta_{N})$ in
 $A_{z}+B_{z}$ and re-calculating the Taylor
series expansion gives:
\begin{eqnarray*}
A_{z}+B_{z} &=& \sum_{i=0}^{\infty} 
    \left(
    1 + 2 \sum_{j=1}^i \cos(2 j \theta_N)
    \right) z^{i}.
\end{eqnarray*}
Since the Taylor series expansion has the form 
\begin{eqnarray*}
A_{z}+B_{z} &=& \sum_{i=0}^{\infty} 
          \left(
          A_{i}+B_{i}\right) z^{i},
\end{eqnarray*}
we conclude that                                                     
\begin{eqnarray*}
A_{i}+B_{i} &=& 
 1 + 2 \sum_{j=1}^i \cos(2 j \theta_N) .
\end{eqnarray*}

The probability of success $P(C,N)$ is $\norm{~ A_{C} + B_{C} ~}^2/N$.
However, it is easy to see from equations (\ref{eq:A}) and (\ref{eq:B})
 that for any positive naturals $C,N$,
$A_{C} + B_{C}$ is real and not complex so that 
$\norm{~ A_{C} + B_{C} ~}^2 =  ( A_{C} + B_{C} )^2$.
We then obtain a closed form for the probability:
\begin{eqnarray*}
P(N,C) &=& \norm{~ A_{C} + B_{C} ~}^2 / N \\
 &=& (A_C+B_C)^2 / N \\
       &=& \left( 1 + 2 \sum_{j=1}^C \cos(2 j \theta_N) \right)^2 / N \\
       &=& \sin^2((2 C+1) \theta_N).
\end{eqnarray*}

\section*{Acknowledgements}

Thanks to Tony Hey and other members of the Quantum Sticky Bun Club 
for inspiration and to Peter H{\o}yer for comments on a draft of the 
paper.

\end{document}